\documentclass[twocolumn,amssymb,showpacs,aps,prl]{revtex4-1}
\usepackage{}
\usepackage{stmaryrd}
\usepackage{amsmath}
\usepackage{amssymb}
\usepackage{float}
\usepackage{graphicx}
\usepackage{titletoc}
\usepackage{booktabs}
\usepackage{array, color}
\usepackage{tabularx}
\usepackage[colorlinks,linkcolor=blue,anchorcolor=blue,citecolor=blue,urlcolor=blue,dvipdfm]{hyperref}
\usepackage{latexsym,bm}
\usepackage{graphics}

\begin{document}

\title{Dynamic orbitals in high-harmonic generation from CO molecules}

\author{Bin Zhang}
\author{Jianmin Yuan}
\author{Zengxiu Zhao}
\email{zhao.zengxiu@gmail.com}
\affiliation{Department of Physics, College of Science, National University of Defense Technology, Changsha 410073, Hunan, People's Republic of China}

\date{\today}

\begin{abstract}
We identify that both the dynamic core polarization and dynamic orbital deformation are important in the orientation-dependent high-harmonic generation of CO molecules subjected to intense few cycle laser fields.
These polarization dynamics allow for the observation of strong orientation effects and dynamic minimum in the harmonic spectra.
The generated attosecond pulses can be greatly affected by these multielectron effects.
This work sheds light on future development of dynamic orbital imaging on attosecond time scale.
\end{abstract}


\maketitle


High-order harmonic generation (HHG) is a fundamental atomic process in strong laser fields that continues to attract much attention~\cite{brabec00,posthumus04}, especially when the recent advances in laser technology have made possible the ultrafast probing of molecular dynamics with ultra-strong laser intensity~\cite{krausz09,kling08,stapelfeldt03}.
As an efficient source of coherent attosecond XUV radiation~\cite{Rundquist98,Bartels00}, HHG opens a route to observing and controlling the fastest electronic processes in matter.
The intrinsic timing of HHG  associated with the interval between ionization and recombination can, e.g., be used to resolve electron dynamics on attosecond time scale~\cite{smirnova09,Haessler10,Mairesse10,xie12} and reconstruct molecular orbitals with \r{A}ngstr\"{o}m spatial resolution~\cite{Itatani04,Vozzi11,Patchkovskii06,Patchkovskii07} for orbital tomography.
While methods based on single electron approximation are widely used~\cite{Itatani04,Bertrand12,morishita07,awasthi08,samba10,Tong02,kjeldsen05,kfr,lewenstein94}, it has been evidenced that multielectron effects come into play for strong-field physics~\cite{McFarland08,Akagi09,Worner10,zhang13,smirnova09,Haessler10,Mairesse10,Patchkovskii06,zhao07pra}.
Identifying and understanding of the encoded multielectron fingerprints in HHG with a self-coherent theory is the key to the applications of both the attosecond pulse (ASP) generation and high-harmonic tomography.

Recently, the dynamic core polarization has been identified to play an important role in the strong-field tunneling ionization of CO molecules~\cite{zhang13}.
The neglect of core polarization dynamics generally leads to qualitatively incorrect ionization yields as compared to the experiments~\cite{wu12,Holmegaard10}.
For HHG, not only the ionization process, but also the electron dynamics in the subsequent propagation and rescattering steps~\cite{corkum93} may be modified by the core polarization field.
Upon recombination, the electron might see the orbitals deformed~\cite{neidel13,spiewanowski13,spiewanowski14,sukiasyan10} due to the interplay of laser fields and polarization fields from the core electrons.
Thus it's desiring to investigate whether and to what degree the HHG signal is affected by the collective response of the electrons.
On the other hand, previous investigations on harmonic emissions from multielectron molecules are generally focused on the non-polar molecules~\cite{smirnova09,Itatani04,Mairesse10,Haessler10,Vozzi11,Patchkovskii07,Patchkovskii06}.
For HHG from polar molecules, not only the alignment~\cite{stapelfeldt03}, but also the orientation effects come into play, benefiting the tailoring and shaping of ASPs through  their generation.

In this letter, we investigate the orientation-dependent HHG of CO molecules in intense few cycle laser fields by the fully propagated time-dependent Hartree-Fock (TDHF) theory~\cite{kulander87}. TDHF goes beyond the single active electron approach and includes the response of all electrons to the field, which helps to partially resolve the multielectron effects from the molecular core within the framework of Hartree-Fock.
We show that the dynamics of core electrons effectively modify the behaviors of HOMO electrons in two senses: detuning the recursion time of the free electron during ionization and recombination; and more importantly, deforming the highest occupied molecular orbital (HOMO) at the instant of recombination. Therefore both the intensity and emission time of the generated ASP are modified, which could be crucial in ultra-fast experiments since the duration of the ASP  itself is in the order of a few hounded attoseconds. Furthermore, the dynamic orbital deformation allows for the observation of dynamic minimum in the harmonic spectra, which provides the possibility of imaging dynamic orbitals.

For the numerical implementation,
we use the prolate spheroidal coordinates~\cite{mathbook65}, which is the most natural choice for two-center systems.
Our approach is also based on the discrete-variable representation and the finite-elements method~\cite{rescigno00}.
The numerical parameters are as follows.
The internuclear distance of CO is fixed at experimental equilibrium of 2.132 a.u.~\cite{nist}.
As the ground electronic state is $^1\Sigma$, spin-restricted form of TDHF is adopted.
The electronic configuration of the ground state of CO from imaginary time relaxation calculation is $(1\sigma)^2(2\sigma)^2(3\sigma)^2(4\sigma)^2(1\pi)^4(5\sigma)^2$, with the total ($5\sigma$) energy of -112.7909118 (-0.554923304) a.u., in good agreement with literature values~\cite{kobus93}.
The asymmetric electron density distribution results in the non-vanishing permanent dipole moment (DM) of each orbital.

\begin{figure}
\includegraphics*[width=3in]{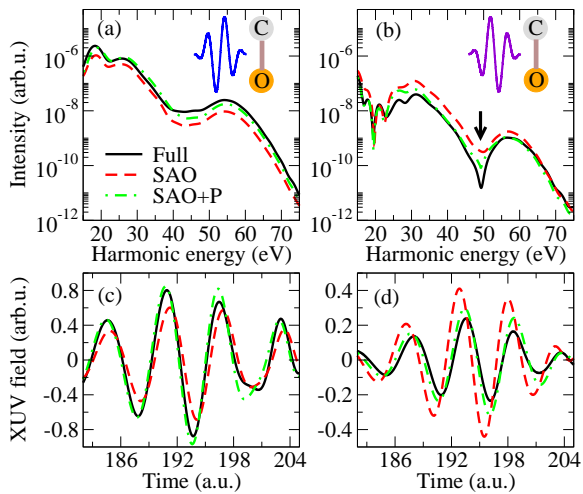}
\caption{(Color online)
Harmonic spectra from $5\sigma$ orbital of CO for:
(a) parallel orientation ($\theta=0^o$); (b) anti-parallel orientation ($\theta=180^o$).
Temporal profiles of the ASP constructed from the 15th to 35th harmonics are presented in (c) for $0^o$ and (d) for $180^o$.
The results are from the full method (solid lines), SAO method (dashed lines) and SAO+P method (dot-dashed lines), respectively.
The black arrow in (b) indicates the minimum position.
The insets visualize the 3 optical cycle sin-squared laser fields, with a carrier wavelength of 800nm.
The intensity corresponding to the maximum electric field is 2$\times 10^{14}$W/cm$^2$.
}\label{fhhgpol}
\end{figure}

The linearly polarized laser fields used in this letter are visualized in the insets of Fig.~\ref{fhhgpol}.
Only the second cycle forms a complete oscillation where both the positive and negative electric fields reach the maximum ($E_{max}=0.0755$a.u.), corresponding to an intensity of 2$\times 10^{14}$W/cm$^2$.
For parallel orientation ($\theta=0^o$) [Fig.~\ref{fhhgpol}(a)], the electron is ionized from and then recombines to the C-end, while ionization and recombination take place toward the O-end for anti-parallel orientation ($\theta=180^o$) [Fig.\ref{fhhgpol}(b)].
Thus, orientation effects can be separated and identified.
Though contributes little to the total harmonic spectra, the first cycle is necessary so that electron dynamics can be included, which helps to the verification of dynamic core polarization with laser-deformed orbitals.
Such single cycle pulses have been enabling the precision control of electron motion and the study of
electron-electron interactions with a resolution approaching the atomic unit of time (24 as) by varying their waveforms~\cite{Goulielmakis08,Marceau13,Hu13}.

The emitted harmonic spectra is given by $S(\omega)=4\omega^4/(6\pi c^3)|\int_0^T {\bf d}(t)e^{i\omega t}\textrm{d}t|^2$, where $c$ is the speed of light, and ${\bf d}(t)=\langle \psi(t)|\hat{\bf r} |\psi(t)\rangle$.
In addition to the full TDHF method, we have also performed the calculations using the single active orbital (SAO) approximation: i.e., propagate the HOMO electrons while freezing the core electrons.
The harmonic spectra from  $5\sigma$ of CO, using the full and SAO methods, are compared in Fig.~\ref{fhhgpol}(a) and (b).
The cutoff position is found at 36th (56eV), in good agreement with the prediction from the Lewenstein model~\cite{lewenstein94},
$N\omega=1.32I_p+3.17U_p$,
where $I_p$ is the ionization potential, $U_p=E_{max}^2/4\omega_L$ is the ponderomotive energy, and $\omega_L$ is the carrier frequency  of the laser field.
Here, $I_p$ is taken as the negative of orbital energy ($\epsilon_{5\sigma}$).
For both orientations, the spectra from SAO method deviate from the results from full method:
SAO method tends to underestimate the full method for $\theta=0^o$, while overestimation is observed for $\theta=180^o$.
This resembles the difference of ionization yields between the full and SAO methods~\cite{zhang13}.
In SAO method the inner orbitals are kept frozen, resulting in an static average potential,
while in SAO+P method the dynamic core polarization is included by a model potential~\cite{zhao07,zhao14},
$V_p({\bf r},t)=-{\boldsymbol{\alpha}_c{\bf E}(t)\cdot {\bf r}}/{r^3}$,
where $\boldsymbol{\alpha}_c$ is the total polarizability of core electrons~\cite{zhang13}.
Close to the core, we apply a cutoff for $V_p$, at a point where the polarization field cancels the laser field~\cite{zhao07}.
This is also necessary to remove the singularity near the core.
By incorporating the dynamic core polarization potential, SAO+P method yields nearly
an all spectra improvement over the original SAO method.
The increase ($\theta=0^o$) and decrease ($\theta=180^o$) of the spectral intensity are originated from the variation of ionization rates due to
core polarization~\cite{zhang13,zhao07}.


How does the core polarization affect the emission time of harmonics, as a consequence of the free electron dynamics?
To this end, we construct the temporal profile of the ASP from the wavelet transform~\cite{Tong00} of  the harmonic spectra
between 15th and 35th, shown in Fig.~\ref{fhhgpol}(c) and (d). For $\theta=0^o$, the ASP from SAO method deviates from the full result both in the intensity and in the emission time with a delay of $\Delta t=t_\textrm{SAO}-t_\textrm{Full}=12$as.
Contrast to $\theta=0^o$ where core polarization advances the emission of ASP, the emission is delayed for $\theta=180^o$($\Delta t=-20$as).
The inclusion of core polarization potential in the SAO+P method improves the results for both orientations.
These different changes by core polarization will be explained later.

\begin{figure}
\includegraphics*[width=3in]{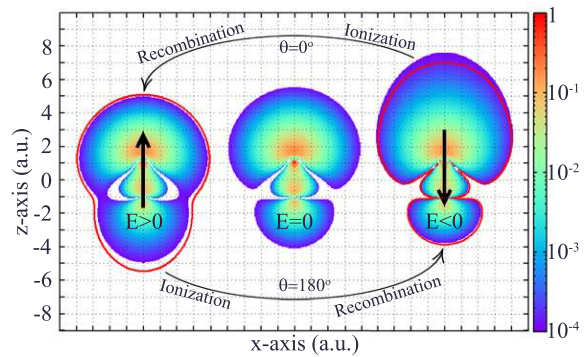}
\caption{(Color online)
The field-free (middle) and laser-deformed orbital densities of $5\sigma$ when the electric field points from O to C ($E>0$, left) and from C to O ($E<0$, right), from full TDHF calculations.
The red lines give the density profiles from SAO calculations.
For the left and right figures, $|E|=E_{max}$.
}\label{fden}
\end{figure}

Strong orientation effects are observed in the harmonic spectra in Fig.~\ref{fhhgpol}: (1) the spectral intensity is generally lower for $\theta=180^o$; (2) an apparent minimum is observed at the 31th (48eV) harmonic for $\theta=180^o$, while no obvious minimum exists for $\theta=0^o$.
These effects already exist in SAO method, which are strengthened in SAO+P and full methods.
In the following we show that these orientation-dependent differences originate from the dynamic deformation of the electronic orbital before recombination.

Figure~\ref{fden} visualizes the field-free and laser-deformed orbital density of $5\sigma$ on x-z plane.
Subjected to strong electric fields, the orbital polarizes and deviates from the field-free distribution, resulting in two distinct orbitals for both cases of the electric fields $E>0$ and $E<0$.
The deformation is stronger for $E<0$, since the major part of the electron density locates at the C-end.
Interestingly, because of the inclusion of core polarization in full calculations, the deformation is enhanced for $E<0$, while suppression is observed for $E>0$.
The incremental changes amount up to $30\%$ of the total orbital deformation in SAO calculations, which further enlarges the density asymmetry between positive and negative electric fields.

According to the molecular tunneling ionization theory~\cite{Tong02,kjeldsen05,madsen12}, ionization rate depends on the density distribution in the asymptotic region ($r\gg 0$).
For $\theta=0^o$ ($180^o$), the electron is ionized for $E<0$ ($E>0$) from the C-end (O-end), where
the asymptotic density is increased (decreased) due to core polarization.
This causes the corresponding changes in the ionization yields as well as in the spectral intensity.
Note that the explanations presented here on ionization consists with the
tunneling picture in our previous work~\cite{zhang13}.
In a recent experiment~\cite{neidel13}, obvious variations were observed in the ionization by an ASP,
due to orbital deformation by a moderately strong near-infrared laser field.
For $\theta=0^o$ ($180^o$), the ionized wavepacket recombines when $E>0$ ($E<0$) towards the C-end (O-end),
thus the recombining wavepacket sees the shrinked (expanded) orbital due to core polarization.
This results in the different modulations in the emission time of ASP.

\begin{figure}
\includegraphics*[width=3in]{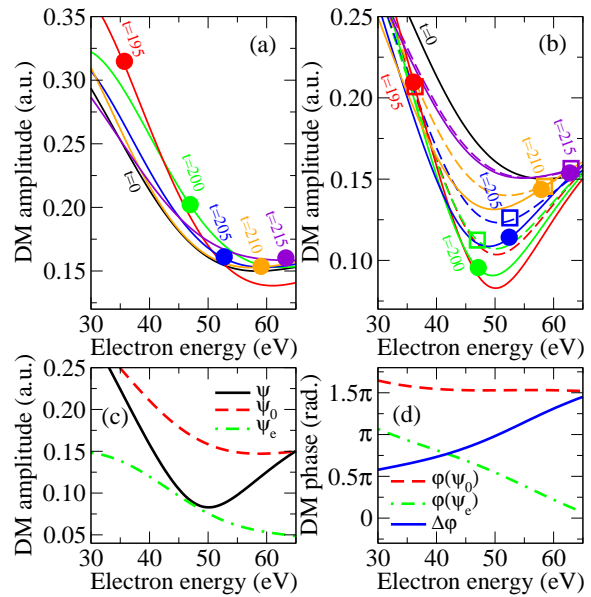}
\caption{(Color online)
(a, b): recombination DMs extracted from $5\sigma$ orbital for (a) $\theta=0^o$ and (b) $\theta=180^o$, at different times.
The solid (dashed) lines are the results from the full (SAO) method.
For each time, the solid circle (open square) marks the corresponding energy of the recombinating electron. The field-free DMs ($t=0$a.u.) are also presented for comparison. The laser parameters are the same as in Fig.~\ref{fhhgpol}.
(c, d): amplitude (c) and phase (d) of recombination DMs extracted from $\psi=\psi_0+\psi_e$, $\psi_0$ and $\psi_e$ (see texts), at $t=195a.u.$ for $\theta=180^o$. The results are from the full method.
}\label{ftdip}
\end{figure}

To further investigate the effects of orbital deformation on harmonic emission,
we calculate the time-dependent photoelectron recombination DM,
$d_z(p_z,t)=\langle p_z |z|\psi(t)\rangle$,
where $|p_z\rangle=(2\pi)^{-3/2}\exp\{ip_z z\}$,
$\psi(t)$ is the time-dependent $5\sigma$ orbital from TDHF calculations.
Fig.~\ref{ftdip} presents the recombination DM versus electron energy ($\epsilon=p_z^2/2$) at different recombination times: 195a.u., 200a.u., 205a.u., 210a.u. and 215a.u.
For comparison, the field-free DM (t=0a.u.) is also presented.
The time-dependent DM deviates from the field-free one mostly when the field reaches $E_{max}$ ($t\approx$ 195 a.u.).
As $t$ increases, the field amplitude decreases and the DM approaches the field-free one (e.g., $t=215$a.u.).
For each line, the filled circle (open square) marks the corresponding energy of the electron at that recombination time.
If the ionized wavepacket does see the deformed orbital upon recombination, the marks should reflect the spectral profiles. Indeed, for $\theta=0^o$, the marks form no clear minimum, in accord with the harmonic spectral profile [Fig.~\ref{fhhgpol}(a)].
And for $\theta=180^o$, the marks exhibit a deep minimum at $\epsilon=47$eV, in good agreement with the position of the minimum in the harmonic spectra [Fig.~\ref{fhhgpol}(b)].
Orbital deformation leads to the suppression of the DMs in (b), while mild enhancement is observed in (a).
This assists the observed suppression of the spectral intensity in Fig.~\ref{fhhgpol}(b).
For SAO calculations, the time-dependent DM deviates smaller from the field-free DM than the full calculation.
This results in the deeper minimum for the spectra from full calculations.

For $\theta=180^o$, the dynamic minimum in harmonic spectra originates from the time-dependent variation of recombination DM.
In the following, we show that at each fixed recombination time, the minimum in the DM comes from the interference
between the field-free and field-polarized components of $5\sigma$ orbital.
To this end, the laser-deformed orbital is written as $\psi(t)=\psi_0(t)+\psi_e(t)$, where the field-free part $\psi_0(t)=c_0(t)\psi(0)$, $c_0(t)=\langle \psi(0)|\psi(t)\rangle$, and the field-polarized part $\psi_e(t)=\psi(t)-\psi_0(t)$.
Fig.~\ref{ftdip}(c) and (d) present the amplitudes and phases of the recombination DMs using these partial waves, at $t=195$a.u.
For both $\psi_0$ and $\psi_e$, $|d_z|$ decreases monotonously, displaying no minimum.
Interference minimum occurs if the phases $\varphi$ of these DMs differ by $\pi$.
Indeed, $\Delta\varphi$ reaches the vicinity of $\pi$ around 50eV, resulting in the minimum in the DM from the total orbital $\psi$.
At other times with smaller $|E|$ and so as smaller orbital deformation ($\psi_e$), the interference minimum decreases due to deviations in the DM amplitudes.

\begin{figure}
\includegraphics*[width=3in]{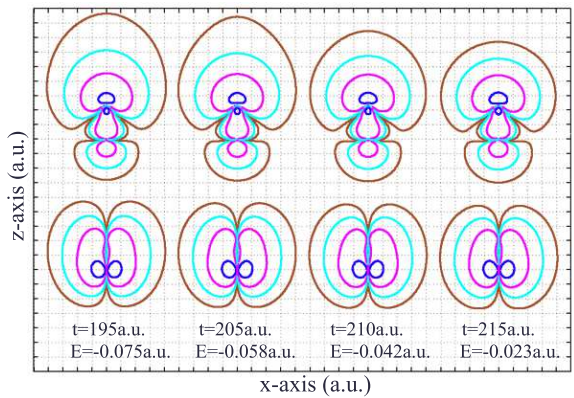}
\caption{(Color online)
Time varying orbital density of $5\sigma$ (upper) and $1\pi$ (lower) computed from full calculations.
From inside, the contour lines are $10^{-1}$, $10^{-2}$, $10^{-3}$, and $10^{-4}$, respectively.
}\label{ftden}
\end{figure}

First thing to note is that no minimum would be found if the electron sees the field-free orbital upon recombination for both orientations.
The existence of deep minimum for $\theta=180^o$ is a clear evidence that the dynamic orbital deformation can greatly affect the process of harmonic generation.
The effects of orbital deformation in harmonic generation was investigated by an extended Lewenstein model~\cite{spiewanowski13}, where
orbital distortion may leads to the disappearance of the two-center inference minimum in harmonic spectra from $N_2$.
The correspondence of harmonics at different orders to different laser-deformed orbitals, underlies the bases for the possible application of dynamic orbital imaging on attosecond time scale.
Fig.~\ref{ftden} presents the laser-deformed orbital density of $5\sigma$ and $1\pi$ at different recombination times.
Only the $E<0$ cases are presented for the illustration, where the deformation is stronger.
The degree of deformation generally follows the amplitude of the laser field.
The density distribution recovers the field free one for small $|E|$.
This agrees with the time-dependent DMs in Fig.~\ref{ftdip}.

The orbital deformations due to core polarization may be explained through the shape of the field-free $5\sigma$ and
$1\pi$ orbitals. At the C-side, the $1\pi$ electrons are distributed more compact comparing to the $5\sigma$ electrons. While at the O-side, quite a portion of $1\pi$ electrons are located outside $5\sigma$. Therefore, for $E<0$, the polarization of $1\pi$ pushes $5\sigma$ further away from the Carbon atom, giving larger polarization of $5\sigma$. On the other hand for $E>0$, the polarization of $1\pi$ in fact suppresses the polarization of $5\sigma$ because of the blocking of $1\pi$ electrons outside the $5\sigma$ at the O-side.

Note that in Fig.~\ref{fhhgpol}, the SAO+P method does not fully reproduce the full method.
There are two possible reasons.
Firstly, the form of polarization potential is accurate in the asymptotic region, but less
valid if the electron approaches the nuclei too closely~\cite{zhao07}.
As a result, the deformation of $5\sigma$ orbital due to dynamic core polarization is not fully incorporated inside the cutoff region.
Secondly, the exchange potential is kept static in the SAO+P method, which in actual fact depends on the dynamic orbital deformation as well.
Despite these facts, the core polarization potential already captures the main sprit that is lost in the original SAO method.
Because the DM directly reflects the electron dynamics, this is a clear evidence that the dynamics of core electrons effectively modify the behaviors of $5\sigma$ electrons. In particular, it reflects that the deformation of the HOMO from the core electrons is essentially an adiabatic process since the probability of nonadiabatic excitation and ionization of the inner-shell electrons is small in the present study.
As a result, the dynamic interaction from core electrons are imprinted in the harmonic spectra emitted from $5\sigma$.

In conclusion, we have identified and demonstrated the importance of the dynamic orbital polarization both for core and HOMO orbitals in the orientation-dependent harmonic spectra.
The polarization dynamics are expected to have more profound effects for molecules with larger polarizability~\cite{neidel13,spiewanowski14}.
The field intensity and emission time of the generated ASP can be greatly affected by these two effects.
Besides the single cycle pulses, the probe of orientation effects in harmonic spectra of hetero-nuclear molecules may also utilize the two-color field scheme, such as the double optical gating~\cite{mashiko08}, where the field itself breaks the up-down symmetry.

\begin{acknowledgments}
This work is supported by  the National Basic Research Program of China (973 Program)
under Grant No. 2013CB922203, the NSF of China  (Grant No. 11374366 and 11274383) and the Major Research plan of NSF of China (Grant No. 91121017).
B. Z. is supported by  the Innovation Foundation of NUDT under Grant No. B110204
and the Hunan Provincial Innovation Foundation For Postgraduate
under Grant No. CX2011B010.
\end{acknowledgments}



\end{document}